\documentclass[twocolumn,aps,prl,showpacs,floatfix]{revtex4}
\usepackage{graphicx}
\usepackage{amsmath}
\usepackage{hyperref}
\usepackage{latexsym}
\usepackage{color}
\usepackage{epsfig}


\begin{document}

\title{Ultracold heteronuclear molecules in a 3D optical lattice}

\author{C. Ospelkaus, S. Ospelkaus, L. Humbert, P. Ernst, K. Sengstock, and K. Bongs}
\affiliation{Institut f\"ur Laserphysik, Luruper Chaussee 149, 22761~Hamburg / Germany}

\begin{abstract}
We report on the creation of ultracold heteronuclear molecules assembled from fermionic $^{40}$K and bosonic $^{87}$Rb atoms in a 3D optical lattice. Molecules are produced at a heteronuclear Feshbach resonance both on the attractive and the repulsive side of the resonance. We precisely determine the binding energy of the heteronuclear molecules from rf spectroscopy across the Feshbach resonance. We characterize the lifetime of the molecular sample as a function of magnetic field and measure between 20 and 120ms. The efficiency of molecule creation via rf association is measured and is found to decrease as expected for more deeply bound molecules.
\end{abstract}


\maketitle

There has been a long quest for production of ultracold molecules in recent years. In particular, heteronuclear molecules would open up intriguing perspectives both in view of their internal properties and their interactions. The electric dipole moment of heteronuclear molecules in their internal ground states makes them one of the best candidates for tests of fundamental physics like the search for a permanent electric dipole moment of the electron and parity violation~\cite{ParViol} as well as for studies on the drifts of fundamental constants. In addition, polar molecules are a key for novel promising quantum computation schemes~\cite{QuantComp}. Furthermore, their large anisotropic interactions give rise to quantum magnetism~\cite{QuantMag}, new types of superfluid pairing~\cite{DipSF} and a variety of quantum phases~\cite{DipPhas}. Currently, two main routes to the production of ultracold molecules are being pursued. One approach aims at cooling thermal ensembles of molecules, e. g. using buffer gas cooling~\cite{BuffCool}, Stark deceleration~\cite{StarkDecel} or velocity filtering~\cite{VelFilt}. The other approach starts with ultracold atomic ensembles and assembles them into molecules by means of photoassociation~\cite{PhotoAsso} or Feshbach resonances~\cite{FeshMol}. In the latter case, one major issue has been the stability of these molecules. While molecules created in bosonic quantum gases have a very short collisional lifetime, bosonic molecules from two fermionic atoms are relatively stable due to the Pauli principle~\cite{FermiFeshMol}. In other cases, as recently demonstrated for bosonic samples \cite{LattMol} and also expected for heteronuclear mixtures, it is favorable to produce the molecules in separated wells of optical lattices to suppress collisional inelastic losses. So far, molecules produced at Feshbach resonances have been limited to homonuclear systems.

In this letter we report on the first creation of ultracold heteronuclear molecules in a 3D optical lattice at a Feshbach resonance. This approach produces ultracold molecules in the ground state of individual lattice sites. This method offers several advantages: long lifetimes allow for further manipulation towards the internal molecular ground state. Moreover, the inherent order within the lattice enables studies of new quantum phases of dipole-dipole interacting systems. In particular, we perform rf association of fermionic $^{40}$K and bosonic $^{87}$Rb atoms close to a heteronuclear Feshbach resonance. Molecules form both on the repulsive and on the attractive side of the resonance, which is a consequence of two-body physics in the presence of strong external confinement. We measure the binding energy as a function of the magnetic field by rf spectroscopy~\cite{rfspect} and study the lifetime and production efficiency of the bound pairs. 

Our experiment has been described previously~\cite{IntDrivDyn,jmo,ImpIndLoc,TunInt}. We create quantum degenerate mixtures of $^{87}$Rb in the $\left|F=2,m_F=2\right>$ state and $^{40}$K in the $\left|F=9/2,m_F=9/2\right>$ state in a crossed optical dipole trap (mean $^{87}$Rb trapping frequency 50 Hz). In order to get a maximum of lattice sites occupied by one boson and one fermion, we have found it beneficial to limit the particle number at this stage to a few ten thousand. We then prepare the mixture in the $^{87}$Rb$\otimes$ $^{40}$K $\left|1,1\right>\otimes\left|9/2,-7/2\right>$ state at a magnetic field close to the Feshbach resonance at 546.8(1) G~\cite{JILAFesh,LENSFesh,TunInt} for the $\left|1,1\right>\otimes\left|9/2,-9/2\right>$ state using the following procedure: First, $^{87}$Rb atoms are transferred from $\left|2,2\right>$ to $\left|1,1\right>$ by a microwave sweep at 20 G and any remaining atoms in the upper hyperfine $\left|2,X\right>$ states are removed by a resonant light pulse. $^{40}$K atoms are transferred into the $\left|9/2,-7/2\right>$ state by performing an rf sweep at the same magnetic field with almost 100\% efficiency. We then increase the magnetic field to a value close to the resonance and ramp up a 3-dimensional optical lattice at a wavelength of 1030nm, where the trapping potential for both species is approximately equal. The optical lattice light is derived from a frequency stabilized 20W Yb:YAG disc laser with a 50ms linewidth of 20kHz. The lattice is formed by three retroreflected laser beams with orthogonal polarizations and a minimum detuning of 10MHz between individual beams.

\begin{figure}[tbp]
\begin{centering}
\leavevmode
\resizebox*{1.0\columnwidth}{!}{\includegraphics{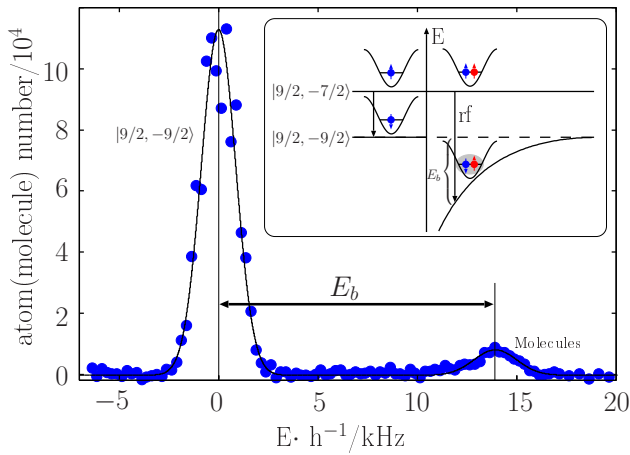}}
\end{centering}
\caption{({\bf color online}) rf spectroscopy of $^{40}$K - $^{87}$Rb in a 3D optical lattice on the $\left|9/2,-7/2\right>\rightarrow\left|9/2,-9/2\right>$ transition (see inset) at a lattice depth of
 $U_{\mathrm{lat}}=27.5E_r$ and a magnetic field of 547.13 G, where the interaction is attractive.
$E_r=\hbar^2 k^2 / 2m_{\mathrm{Rb}}$ is the $^{87}$Rb recoil energy. The spectrum is plotted as a function of detuning from the undisturbed atomic resonance frequency and clearly shows the large atomic peak at zero detuning. The peak at 13.9kHz is due to association of $\left|1,1\right>\otimes\left|9/2,-7/2\right>$ atom pairs into a bound state.
}\label{FigRFSpect}
\end{figure}

In the optical lattice, we create molecules by rf association of pairs of one $^{87}$Rb and one $^{40}$K atom at a single lattice site \footnote{We have also created Feshbach molecules by sweeping the magnetic field across the resonance and probed them via rf dissociation. However -- due to limitations in the magnetic field settling time -- for an accurate measurement of the binding energy it is more favorable in our setup to apply the rf association method.} (see Fig.~\ref{FigRFSpect}). The binding energy is measured as a frequency shift between the molecular and atomic feature. In the following, we discuss the measurement procedure in detail. The $\left|1,1\right>\otimes\left|9/2,-7/2\right>$ state that we prepare in the optical lattice has scattering properties which are to a good approximation independent of the magnetic field in the region where we probe the system. The corresponding energy thus serves as reference level for the following measurements. In order to measure interaction shifts and determine molecular binding energies of the Feshbach-resonant $\left|1,1\right>\otimes\left|9/2,-9/2\right>$ mixture, we perform rf spectroscopy on the $^{40}$K $\left|9/2,-7/2\right>\rightarrow\left|9/2,-9/2\right>$ transition in the optical lattice (see inset Fig.~\ref{FigRFSpect}). At magnetic field values close to the Feshbach resonance, this transition occurs at about 80 MHz and has a magnetic field sensitivity of 67 kHz/G. We detect left-over atoms in the $\left|9/2,-7/2\right>$ state at high magnetic field by detuning the imaging laser by -834 MHz with respect to the low-field $\left|9/2,9/2\right>\rightarrow\left|11/2,11/2\right>$ cycling transition. Atoms in the $\left|9/2,-9/2\right>$ state and molecules are probed at a detuning of -765 MHz. Using high field imaging and separate CCD cameras, we can thus state-selectively detect both atoms and weakly bound molecules in a single experimental run. The rf spectroscopy is performed by shining in an rf pulse with a gaussian amplitude envelope with a $1/e^2$ full width of 400 $\mu$s and a total pulse length of 800 $\mu$s, resulting in an rf $1/e^2$ half linewidth of 1.7 kHz. The pulse power is chosen to achieve full transfer on the single atom transition.

Fig.~\ref{FigRFSpect} shows a typical rf spectrum with atomic and molecular resonance peaks. The peak at zero frequency with respect to the unperturbed atomic transition frequency between the two Zeeman substates is the single atom peak from lattice sites occupied by only one fermion. On the high energy side of the spectrum at a detuning of 13.9 kHz with respect to the atomic transition, we observe a distinct feature which is due to rf association of atoms into molecules. The spectrum in Fig.~\ref{FigRFSpect} was recorded at 547.02 G on the high field side of the resonance, where the interaction between bosons and fermions is attractive and stable molecules do not exist in free space. The presence of the optical lattice nevertheless admits a bound state at $a_{\mathrm{FB}}<0$~\cite{ETHMol3DLat}. From the rf spectrum, we can determine the separation between the single atom and the molecular peak with high precision and thus extract the binding energy  up to a constant offset due to non-zero background scattering lengths. At the same time, the atomic peak provides us with a precise magnetic field calibration across the whole resonance. We find that the magnetic field deviates by no more than 9 mG from the calibration over one week, which corresponds to a maximum frequency shift of about 500 Hz.

\begin{figure}[tbp]
\begin{centering}
\leavevmode
\resizebox*{1.0\columnwidth}{!}{\includegraphics{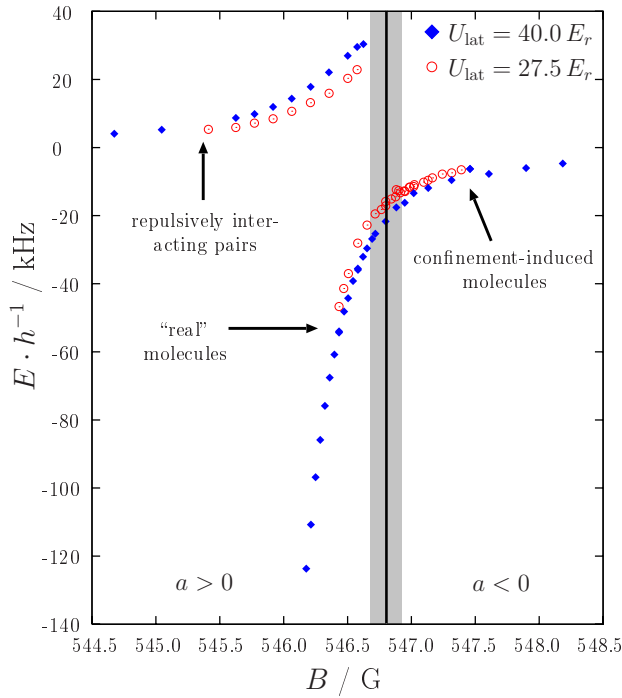}}
\end{centering}
\caption{({\bf color online}) Binding energy of heteronuclear $^{40}$K - $^{87}$Rb molecules in an optical lattice for two different lattice depths $U_{\mathrm{lat}}$ in units of the $^{87}$Rb recoil energy
$E_r=\hbar^2 k^2 / 2m_{\mathrm{Rb}}$. The center of the Feshbach resonance is located at 546.8(1) G~\cite{TunInt}. We observe attractively bound molecules which are confinement-induced at a positive detuning with respect to the resonance center and ``real'' molecules which are stable in free space below the center of the resonance. In addition, we observe repulsively interacting pairs with a positive``binding energy'' below the resonance.
}\label{FigBindEn}
\end{figure}

Fig.~\ref{FigBindEn} as the main result of this paper shows our results for the binding energy as a function of magnetic field across the resonance for two different lattice depths of $40 E_R$ and $27.5 E_r$, respectively, where $E_r=\hbar^2 k^2/2m_{\mathrm{Rb}}$ is the $^{87}$Rb recoil energy \footnote{The lattice depth has been calibrated by observing the response of a $^{87}$Rb cloud to modulation of the lattice potential.}. In free space, molecules close to a Feshbach resonance only exist for $a>0$. This picture is modified in the presence of the external optical potential~\cite{PseudoPotBusch}. In this case, bound states for $a<0$ also exist (so-called confinement-induced molecules~\cite{ETHMol3DLat}). These exhibit a smooth transition into ``real'' molecules bound even in the absence of the external confinement on the $a>0$ side of the Feshbach resonance. This is clearly visible in the bottom branch of the spectrum in Fig.~\ref{FigBindEn}. As the scattering length becomes smaller and smaller again on the repulsive side of the resonance, molecules become more and more deeply bound. On this side of the resonance, we can also excite an atomic pair branch where repulsion between Bosons and Fermions shifts the two-particle ground state towards a higher energy~\cite{PseudoPotBusch}. As expected, Fig.~\ref{FigBindEn} shows that increasing lattice depth tends to increase both the ``positive binding energy'' of the repulsive pairs and the binding energy of the molecular branch away from the noninteracting case.   

\begin{figure}[tbp]
\begin{centering}
\leavevmode
\resizebox*{1.0\columnwidth}{!}{\includegraphics{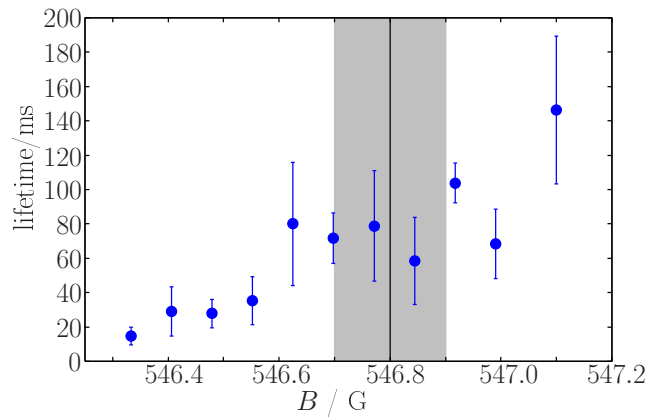}}
\end{centering}
\caption{({\bf color online}) Lifetime of heteronuclear molecules in the optical lattice as a function of magnetic field. On the repulsive side of the resonance below 546.8(1) G the lifetime tends to decrease. 
}\label{FigLifetime}
\end{figure}

One of the main concerns in ultracold Feshbach chemistry is the lifetime of the molecules. In general, these molecules are very fragile objects due to their highly excited internal state. Initial experiments with molecules created from bosonic atoms~\cite{FeshMol} accordingly showed a very short lifetime, which has recently been overcome in the presence of a deep 3D optical lattice~\cite{LattMol}. Molecules created from two fermionic atoms have shown a long lifetime close to Feshbach resonances due to Pauli-forbidden inelastic decay~\cite{PetrovDimers}. For heteronuclear molecules composed of a bosonic and a fermionic atom, the situation is somewhat more complicated; suppression of collisions due to their fermionic character is expected for more deeply bound molecules~\cite{MITFesh}. However, in addition, in this case, remaining free atoms are expected to strongly limit the lifetime of the molecules. We therefore produce the heteronuclear molecules within 3D optical lattices where the large tunneling time strongly reduces inelastic collisional losses. Fig.~\ref{FigLifetime} shows a measurement of the lifetime of the heteronuclear molecular sample in an optical lattice with a depth of 40.0 $E_r$ as a function of magnetic field. We find a lifetime of about 120ms for weakly bound confinement-induced molecules. In the vicinity of the resonance, the lifetime is about 80ms, and drops to 20 to 40ms for more deeply bound molecules. The measurement is performed by rf associating atoms in the $\left|1,1\right>\otimes\left|9/2,-7/2\right>$ state into molecules and transferring any remaining $\left|9/2,-7/2\right>$ atoms into the $\left|9/2,-5/2\right>$ state using an rf sweep at 85 MHz. A resonant light pulse on the $\left|9/2,-7/2\right>$ detection frequency ensures that no atoms in this state are left. Since high-field imaging detects both atoms and weakly bound molecules, it is of potential concern that molecules may fall apart and form unpaired $\left|9/2,-9/2\right>$ atoms. We therefore drive a $\pi$ pulse on the atomic $\left|9/2,-7/2\right>\rightarrow\left|9/2,-9/2\right>$ transition prior to imaging in order to detect any free $\left|9/2,-9/2\right>$ atoms on the $\left|9/2,-7/2\right>$ detection transition, but have found no atomic signal. This proves that our lifetime measurement really measures the lifetime of the molecular sample. The lifetime may be currently limited due to remaining fermionic $^{40}$K atoms which, for this measurement, are in the $\left|9/2,-5/2\right>$ state \footnote{We performed other measurements on the system where $^{40}$K remained in the $\left|9/2,-7/2\right>$ state and found comparable lifetimes.}. For these atoms, the optical lattice potential is only 20 $E_r^{\mathrm{K}}$ deep ($E_r^{\mathrm{K}}=\hbar^2k^2/(2m_{\mathrm{K}})$), which corresponds to a tunneling time on the order of 10 ms. The observed dependency of the lifetime on the binding energy is still an open question.

\begin{figure}[tbp]
\begin{centering}
\leavevmode
\resizebox*{1.0\columnwidth}{!}{\includegraphics{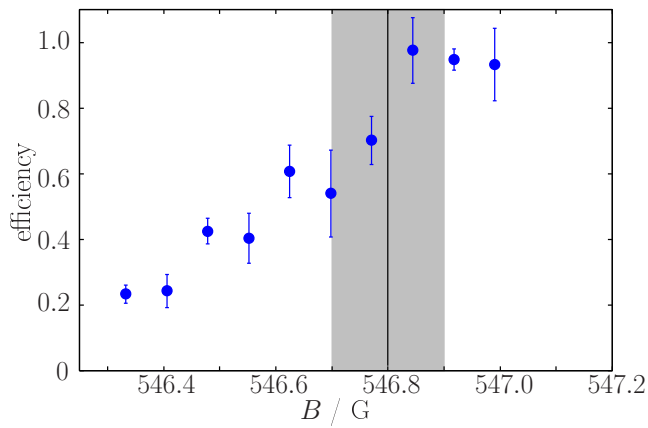}}
\end{centering}
\caption{({\bf color online}) Transfer efficiency of the rf association process as a function of magnetic field. The transfer efficiency is normalized to its peak value on the attractive side of the Feshbach resonance, where we have observed the highest efficiency within the plotted magnetic field range.
}\label{FigTransEff}
\end{figure}

As a further point, we have analyzed the transfer efficiency of the rf association (Fig.~\ref{FigTransEff}). We detect the molecule number as a function of the magnetic field across the Feshbach resonance and normalize it to the high-field side of the resonance, where the overlap between the initial attractively interacting state and the bound state wave function is highest in the magnetic field range covered by our experiments. The transfer efficiency is observed to be almost constant for attractive interactions and then decreases on the repulsive side of the resonance. As expected from the decreasing overlap of atomic and molecular wave functions for more deeply bound molecules, the transfer efficiency drops to about 20\% on the low-field side of the resonance. On the attractive side of the resonance, we achieve a total efficiency for the transfer of $^{40}$K $\left|9/2,-7/2\right>$ into molecular $^{87}$Rb-$^{40}$K states of about 10\%, producing approximately $10^4$ molecules. This corresponds roughly to the estimated fraction of lattice sites with one fermion and one boson as compared to the total fermion number. Purification schemes for the bosonic component and adjustment of the relative population are possible routes to increase the efficiency.

In conclusion, we have for the first time created fermionic heteronuclear Feshbach molecules from a Fermi-Bose mixture in a 3D optical lattice. We have performed rf spectroscopy of the mixture in the optical lattice and measured the binding energy of the molecular sample. We have found confinement-induced pairs, molecules which are bound in the absence of the lattice and repulsive Feshbach pairs for different strength of the lattice confinement over a wide range of Feshbach binding energies. We have studied the lifetime of the molecular sample and found timescales between 20ms for more deeply bound molecules and 120ms on the attractive side of the resonance. We have studied the transfer efficiency of the rf association process across the Feshbach resonance. 

These experiments pave the way towards physics with ultracold heteronuclear molecules. An important next step will be to apply schemes to transfer the weakly bound molecules into polar ground state molecules in 3D lattices. This would open many new perspectives for studies on fundamental constants and their possible time variation as well as for the study of systems with long range anisotropic interactions.

Finally we note that evidence for short-lived $^{40}$K-$^{87}$Rb molecule formation in an optical dipole trap was recently reported by the group of D.~S. Jin at JILA~\cite{JILAHetNucMol}

\begin{acknowledgments}
We acknowledge fruitful discussions with M. Cramer, F. Deuretzbacher, P. Julienne, E.~G.~M. v. Kempen, S.~J.~J.~M.~F. Kokkelmans, D. Pfannkuche, K. Plassmeier, K. Rz{\c a}{\.z}ewski, E. Tiemann, and experimental assistance by J. Kronj\"ager, M. Succo and O. Wille as well as financial support by Deutsche Forschungsgemeinschaft (SPP 1116).
\end{acknowledgments}

\end{document}